# Centaurs and Scattered Disk Objects in the Thermal Infrared: Analysis of WISE/NEOWISE Observations

James M. Bauer[1,2], Tommy Grav[3], Erin Blauvelt[1], A. K. Mainzer[1], Joseph R. Masiero[1], Rachel Stevenson[1], Emily Kramer[1,4], Yan R. Fernández[4], C. M. Lisse[5], Roc M. Cutri[2], Paul R. Weissman[1], John W. Dailey[2], Frank J. Masci[2], Russel Walker[6], Adam Waszczak[7], Carrie R. Nugent[1], Karen J. Meech[8,9], Andrew Lucas[2], George Pearman[2,10], Ashlee Wilkins[1,11], Jessica Watkins[12], Shrinivas Kulkarni[13], Edward L. Wright[14], and the WISE and PTF Teams

[1]Jet Propulsion Laboratory, California Institute of Technology, 4800 Oak Grove Drive, MS 183-401, Pasadena, CA 91109 (email: bauer@scn.jpl.nasa.gov)

[2]Infrared Processing and Analysis Center, California Institute of Technology, Pasadena, CA 91125

[3]Planetary Science Institute, 1700 East Fort Lowell, Suite 106, Tucson, AZ 85719-2395

[4]Department of Physics, University of Central Florida, 4000 Central Florida Blvd., P.S. Building, Orlando, FL 32816-2385

[5] Applied Physics Laboratory, Johns Hopkins University, 11100 Johns Hopkins Road Laurel, MD 20723---6099

[6] Monterey Institute for Research in Astronomy, 200 Eighth Street, Marina, CA 93933

[7] Division of Geological and Planetary Sciences, California Institute of Technology, Pasadena, CA 91125

[8]Institute for Astronomy, University of Hawaii, 2680 Woodlawn Dr., Manoa, HI 96822

[9]NASA Astrobiology Institute, Institute for Astronomy, University of Hawaii, Manoa, HI 96822

[10] University of California, Berkeley

[11]Dept. of Astronomy, Univ. of Maryland, College Park, MD 20742-2421

[12]Institute for Planets and Exoplanets, University of California, 3713 Geology, 595 Charles Young Drive East, Los Angeles, CA 90095

[13] Division of Mathematics, Physics, and Astronomy, California Institute of Technology, Pasadena, CA 91125

[14]Department of Physics and Astronomy, University of California, PO Box 91547, Los Angeles, CA 90095-1547

**ABSTRACT**

The Wide-field Infrared Survey Explorer (WISE) observed 52 Centaurs and Scattered Disk Objects in the thermal infrared, including the 15 discoveries that were new. We present analyses of these observations to estimate sizes and mean optical albedos. We find mean albedos of 0.08 +/- 0.04 for the entire data set. Thermal fits yield average beaming parameters of 0.9 +/- 0.2 that are similar for both SDO and Centaur sub-classes. Biased cumulative size distributions yield size-frequency distribution power law indices ~ -1.7 +/- 0.3. The data also reveal a relation between albedo and color at the 3-sigma level. No significant relation between diameter and albedos is found.

## 1. Introduction

The region of the outer solar system between the giant planets is a dynamically unstable domain for small body orbits. The cometary and asteroidal bodies that spend the majority of their time in this region, the Centaurs, typically have dynamical lifetimes of only tens of millions of years (cf. Holman & Wisdom, 1993 and Horner et al. 2004). Centaurs form a link between the more distant populations, such as the Scattered Disk Objects (SDOs) and Trans-Neptunian Objects (TNOs) with populations that reside closer in towards the Sun, such as the Jupiter Family Comets (JFCs) and Near Earth Comets (NECs).

As the transition stage between the TNO reservoir of bodies (Gladman et al. 2008) that evolve to become inner system JFCs, Centaurs (Horner et al. 2004) and SDOs (Gladman et al. 2008; Elliot et al. 2005) serve as a means of transport of volatiles to the terrestrial planets and inner solar system. The diversity of observable properties in the JFC population (A'Hearn et al. 1995, Schleicher et al. 2012, Fernández et al. 2013, Kelley et al. 2013) and their variances in behavior and gas species abundances (Bauer et al. 2011 & 2012a,b,c) may be demonstrative of evolutionary processes, or may be intrinsic properties related to their formation distances.

Whether evolutionary effects are manifest in the reservoir populations (Brown 2000, Lamy et al. 2001, Jewitt 2004) or whether the differences in the observable surface properties of these populations, for example in color (Tegler et al. 2008 and references therein), are linked to primordial compositional variations has not been definitively determined. The albedo ($p_v$) of JFCs, 0.04 +/- 0.02 (Fernández et al. 2013), is on average less than the observed albedos for SDOs and Centaurs (as detailed below). Should color be linked to albedo, these properties may trace the surface volatile content, and thus the age and evolution of these bodies (cf. Jewitt 2009, Fraser & Brown 2012, Lisse et al. 2013).

Numerous schemes have been proposed for the classification and nomenclature of outer solar system bodies (cf. Horner et al. 2004, Elliot et al. 2005, Delsanti & Jewitt 2006, Gladman et al. 2008). All are primarily founded on arguments concerning the dynamical evolution of the related objects. However, owing to the distance of these bodies, the pace of their physical characterization has been ponderous. In most cases, large-aperture telescopes are necessary to make accurate photometric and spectro-photometric observations (c.f. Bauer et al. 2003; Delsanti et al. 2006; Tegler et al. 2008). Even the most basic physical property, size, has remained elusive because the surface-reflectances and colors of these objects can vary by factors of several (Jewitt and Kalas 1998, Fernández et al. 2002, Barucci et al. 2008), leading to sizes based on reflected light having commensurately large uncertainties. With the advent of space-based thermal infrared observations with the Spitzer Space Telescope (SST; Werner et al. 2004), the derivation of sizes for many of these bodies became possible. Prior to SST, only a handful of outer solar system bodies had measured diameters in the literature (Fernández et al. 2002). The work of Stansberry et al. (2008) demonstrated the effectiveness of such thermal IR observations by providing diameters for 47 of these objects, and setting constraints on surface reflectance and thermal beaming parameters, a value that indicates the degree that the small body‘s emission deviates from that of an idealized sphere in instantaneous thermal equilibrium (Harris 1998). The data set was large enough to make comparisons that were statistically meaningful with respect to dynamical sub-classes, but indicated at best only weak relationships between the subclasses. Recently, the Herschel spacecraft, too, has sampled 15 SDOs (Santos-Sanz et al. 2012, hereafter S12). Both S08 and S12 have some overlap with the WISE sample of objects.

The WISE mission surveyed the entire sky at four mid-IR wavelengths simultaneously, 3.4 μm (W1), 4.6 μm (W2), 12 μm (W3) and 22 μm (W4), with greatly improved sensitivity and resolution over previous surveys at these wavelengths (Wright et al. 2010). The field of view for each WISE exposure was 47 arcmin on a side, with a pixel scale of 2.75 arcseconds/pixel. Each point on the sky was observed an average of 12 times, with depth of coverage increasing towards the ecliptic poles (Cutri et al. 2012). For this paper we have utilized the NEOWISE enhancements to the WISE data processing system (Mainzer et al. 2011), that provided a solar system object archive and a moving object detection algorithm. The WISE Moving Object Processing Software (WMOPS; Dailey *et al.* 2010; Mainzer *et al.* 2011a) successfully found a wide array of primitive bodies detected by WISE, including Near-Earth Objects (NEOs), main belt asteroids, comets, and Trojan and Centaur asteroids. By the end of the spacecraft mission, NEOWISE identified more than 158,000 small bodies, including 155 comets (cf. Mainzer *et al.* 2011a, Bauer *et al.* 2012c) and 157,000 asteroids (Masiero et al. 2011). The archive augmentation identifies all the WISE observations that covered the predicted positions of moving objects. Many of the Centaurs and SDOs were found over the course of the mission by WMOPS, but a large fraction of the fainter Centaurs were found by stacking the images that were identified as covering the particular Centaur in question using the archive service.

For asteroids and cometary nuclei, the WISE observations were found to be useful for determining solid body size and albedo distributions, and thermo-physical properties such as thermal inertia, the magnitude of non-gravitational forces, and surface roughness (Mainzer *et al.* 2011b, 2011c, Bauer et al. 2012a, and Nugent at al. 2013). The intended scope of this paper is to present the thermal fits of the Centaurs and SDOs and their immediate interpretation. The sample is not de-biased. Many of the detections of the bodies were from stacked images, and sensitivity was greatly affected by object distance and size, rate of apparent motion, and surface reflectance.  For the objects detected through stacking, their selection biases depend on the circumstances of their discovery, usually by ground-based surveys searching at visible wavelengths.  Such surveys are intrinsically biased against lower albedo objects.  Thus, the combined survey biases of the Centaur and SDO sample in this paper consist of a mix of the infrared/WMOPS-selected sample, as well as the visual-wavelength-selected sample detected through stacking.  De-biasing the observed population, and the interpretation of the de-biasing with respect to the underlying populations, will be covered in a paper soon to follow.

All objects we discuss reside primarily within or cross through the region of the giant planet orbits and have semi-major axes, *a*, beyond the orbit of Jupiter, so that in the broadest definition (cf. Levinson and Duncan 1997) they may be considered Centaurs. A similar definition for Centaurs is employed by the JPL solar system dynamics node (http://ssd.jpl.nasa.gov) for all objects with 5.5 AU < *a* < 30.1 AU. We do not impose nomenclature on our subsets, and list the orbital elements (eccentricity, *e*, semi-major axis, *a* [AU], and inclination, *i* [°]) for the reader's interpretation (See Table 4). However, the definition of a Centaur is not rigidly defined (Davies et al. 2008; Horner et al. 2004), and other combinations of orbital parameters have been used to define or sub-divide this class of objects (Gladman et al. 2008), including that Centaurs be required to have a *q* > 5.2 AU (cf. Tegler et al. 2008; Jedicke et al. 2002). For the purposes of our discussion, we do utilize a working definition, without formal explanation, of a subclass of Centaur objects with perihelion distance, *q* < 5.2AU, and *a* > 5.5 AU, calling them "Satyrs", in order to differentiate between the two classification schemes above. Those objects with aphelion distance Q > 35 AU are categorized as Scattered Disk Objects (SDOs), as they satisfy the criterion described in Elliot et al. (2005) with $T_N < 3$ (Table 4) identified with the population of scattered objects indicated in Duncan & Levison (1997).

## 2. Observations & Analysis

### *2.1 Shorter Wavelength Observations*

The WISE thermal observations provide an effective means of determining the size and beaming parameters for small bodies in general, and the Centaurs and SDOs in our observed sample (see Section 2.2). However, the reflected-light signal for most bodies was completely unconstrained by the WISE observations. For all 52 objects observed, the mean of the object heliocentric distances ($R_h$) at the time of observation was 10.5 AU, and only 10 of these were detected at $R_h < 5$ AU, while only one was observed at $R_h < 2$ AU.

The strength of the signal in each band is determined by a number of factors including $R_h$, phase angle, diameter, surface temperature, surface roughness, rotation rate, and surface reflectance. Since many of these factors were not previously measured, it was difficult to predict with certainty which bodies would have detectable flux in specific bands. A general relationship between $R_h$ (in AU) and the wavelength of peak emission, $\lambda_{peak}$ (in μm), can be derived by using Wien's law and assuming the sub-solar temperature, $T_{ss} \sim 394\ K \bullet \sqrt{R_h}$, dominates the signal, such that:

$$R_h \approx \left(\frac{\lambda_{peak}}{7.35}\right)^2 \quad (1).$$

In general, using this expression for a slowly-rotating small body with low surface-reflectance, the sub-solar temperature would produce a peak flux at a wavelength longward of the W2 bandpass when the object is at a distance of $R_h \sim 0.6$ AU. Similarly, the peak flux would fall longward of W3 at $R_h > 4$ AU, and longward of W4 at $R_h > 10$ AU. Using a more sophisticated thermal model (Section 2.2) in combination with the expected reflected-light signal, the thermal signal would exceed the reflected-light out to distances of $R_h \sim 4$ AU for W2, and $R_h \sim 15$ AU for W3. Folding in the estimated signal-to-noise in each band as reported in Wright et al. 2010, the SNR would be greatest in W2 out to $R_h \sim 1$ AU, and in W3 from $R_h \sim 1$ to 2.3 AU. Beyond $R_h \sim 2.3$ AU, the SNR would be greatest in W4, with the idealized case of $p_V \sim 0.04$ and a slowly-rotating body. Only five out of the 52 outer solar system small bodies detected by WISE had significant signals in W1 or W2, even after stacking: Chiron, 2010 $OR_1$, 2010 $LG_{61}$, 2008 $SO_{218}$, and 2008 $YB_3$ (see Table 1). Owing to possible compositional variation in these bodies, W1 and W2 are not readily convertible to reflected-light V-band magnitudes. We therefore utilized our visual-band measurements and visual-band measurements from the literature in our fits to constrain $p_V$. All objects had estimates of absolute magnitude, $H_v(1,1,0)$, based on discovery and astrometric observation reports provided by the Minor Planet Center (MPC; http://www.minorplatencenter.net). Errors on the H magnitudes were taken to be ±0.3 mag in cases where only MPC data was available following Mainzer et al. 2011a and Masiero et al. 2011; it should be noted that considerable uncertainty about the catalog H and phase curve slope parameter G values exists (c.f. Pravec et al. 2012). However, in order to classify a significant sample as to their Centaur color class (Tegler et al. 2008), we observed a subset of our Centaur/SDO targets from ground-based telescopes (Table 2) and used photometry from the literature where otherwise available (Table 3). A total of 33 Centaurs and SDOs of the 52 WISE objects observed have colors measured in this fashion. We also augmented our sample with eight bodies with measured diameters and color photometry from Stansberry et al. (2008; hereafter S08), so that our total sample for comparative analysis included 60 objects. Table 3 lists the references of known colors and B-R values used in our data analysis.

Over the course of the WISE mission, and the three observing semesters immediately following, the NEOWISE team used ground-based telescopes to obtain visual-band and NIR observations of small-body targets. Time at the SOAR 4-meter and Palomar 200-inch telescopes was used to obtain B and R Bessel-filter photometry (Bessel 1990) of the Centaurs and SDOs in our sample. We list the observed R-band magnitude, and B-R values found from our observations in Table 2, as well as the dates and site information. Most of these objects do not have well-constrained phase curve behavior. We also bracketed the B-band observations by R-band observations, but we found no significant R-band variation owing to rotation or cometary activity over the course of our observations. In most cases, the ground-based exposures spanned less than 40 minutes in total for each object.

*2.2 Thermal Observations*

The WISE spacecraft surveyed the entire sky as its terminator-following polar orbit around the Earth progressed about 1 degree of ecliptic longitude per day. Regular survey operations commenced on Jan. 14, 2010 (MJD 55210), imaging the sky simultaneously in all four photometric bands until the solid hydrogen cryogen was depleted in the secondary tank on Aug. 6, 2010 UTC (MJD 55414). The survey then entered a three-band (W1-W3) phase that lasted through Sep. 29, 2010 UTC (MJD 55469). The final phase, the post-cryogenic mission with only W1 and W2 operating, lasted from through Feb 1, 2011 (MJD 55593; cf. Cutri *et al.* 2012). All photometric data of detected objects presented here were obtained during the cryogenic phase.

During the fully cryogenic portion of the mission, simultaneous exposures in the four WISE wavelength bands were taken once every 11 sec, with exposure durations of 8.8 sec in W3 and W4, and 7.7 sec in W1 and W2 (Wright *et al.* 2010). The number of exposures acquired on each moving object varied depending on the location of the object on the sky, especially its ecliptic latitude, the toggle pattern of the survey employed to avoid imaging the Moon, and the relative motion of the object with respect to the progression of the survey (Mainzer *et al.* 2011a, Cutri *et al.* 2012). Note that WISE may have observed a moving solar system object while it was in different patches of the sky, i.e. when several weeks or months had passed since the previous exposure (e.g., comet 67P; Bauer et al. 2012b); henceforth, we refer to the series of exposures containing the object in the same region of sky as a "visit". The spatial resolution in the WISE images varies with the wavelength of the band. The full-width-half-maximum (FWHM) of the mean point-spread-function (PSF), in units of arcseconds was 6.1, 6.4, 6.5 and 12.0 for W1, W2, W3, and W4, respectively (Wright *et al.* 2010, Cutri et al. 2012).

Like the comets we have previously studied (Bauer et al. 2011, 2012a,b), the analysis of the Centaurs and SDOs often required stacking to obtain reliable signal-to-noise ratios (SNRs). For each body, the images were identified using the WISE image server (http://irsa.ipac.caltech.edu/applications/wise), as described by Cutri et al. (2012), and were stacked using the moving object routine, "A WISE Astronomical Image Co-adder" (AWAIC; Masci & Fowler 2009). The images were stacked to both boost the SNR in the thermal bands and to average over any rotational variations in the signals. The ~10-12 observations per object span ~36 hours for most objects, which is probably a reasonable averaging over most Centaur rotational periods, which are shorter than 27 hours (cf. Rousselot et al. 2005) and in most cases < 11 hours (cf. Bauer et al. 2002 & 2003, Ortiz et al. 2002 & 2003, Thirouin et al. 2010). As the objects were not generally observed at heliocentric distances ($R_h$) of less than 4AU, most objects were brightest in W4, and only four had any unambiguous reflected-light signal at shorter wavelengths (W1 and W2; see Section 1). Figure 1 shows the distribution of the Centaurs and SDOs as a function of $R_h$ and indicates a paucity of small-diameter objects at large heliocentric distances, a biasing artifact of the spacecraft's sensitivity limits in the thermal bands (particularly W4), as well as the visual-wavelength discovery selection bias for the objects recovered in the stacked images. Table 1 shows the extracted flux values for the stacked images and viewing geometries, including phase angles.

We used the thermal model developed specifically for solar system objects in the WISE data described in Mainzer et al. (2011b). This model is an adaptation of the Near-Earth asteroid thermal model (NEATM; Harris 1998). Counts are converted to in-band magnitudes and uncertainties using aperture photometry count values from stacked images and applying the W1-4 zero-point offsets in Wright et al. (2010). The magnitudes require color-corrections based on each object's temperature when converting them into flux densities. These corrections are typically small for W4, on the order of a few percent. For W3, however, these can be several tens of percent, though the corrections are well-characterized (Mainzer et al. 2011b, Wright et al. 2010), yielding flux uncertainties on the order of 5 to 10%. As discussed in Mainzer et al. (2011b), the thermal model converts raw instrumental magnitudes into monochromatic flux densities using the color corrections outlined in detail in Wright et al. (2010), deriving an effective temperature for this conversion based on the object's sub-solar temperature. Error bars on the modeled diameters and visual albedos were determined for each object by running 20 Monte Carlo trials that varied the objects' *H* value and the *WISE* magnitudes within their measurement uncertainties using Gaussian probability distributions. The orbital elements *a, e,* and *i*, and the NEATM fit results are shown in Table 4 for the WISE-observed Centaurs and SDOs. Table 4 also lists the additional 8 objects from S08 that were not covered by the WISE observations, but have measured B-R colors. The fifth and last columns indicate special details of the thermal model fits and processing. The last column indicates whether the object was discovered by WISE (indicated by "D" in the last column), was detected in the WMOPS processing (indicated by "Y"), or was found in subsequent stacking ("N"). Note that an "X" in the last column refers to elements from the S08 results. The entries are listed in categories of Centaurs, in order of perihelion distance *q*, and SDOs, in order of aphelion distance *Q*. The first 14 entries in the Table are the previously mentioned Satyr objects. The "NEATM fits comment" column indicates whether a fixed beaming parameter $\eta$ was used for the conditions of the thermal fit. In most cases where there was significant W3 and W4 signal, $\eta$ was fit as a free parameter in the modified NEATM model as implemented using the same method described in Grav et al. (2011, 2012a,b), Mainzer et al. (2011b,c), and Masiero et al. (2011). Cases for which $\eta$ could be fit as a free parameter are labeled "Free" in the fit comments column in Table 4. However, in cases where the freely-varying $\eta$ fit returned non-physical values, i.e. $\eta < 0.5$ (Lebofsky & Spencer 1989, Harris 1998), a fixed $\eta$ value of 0.8 was used if there were no NEATM fits in S08 or S12 with freely fit $\eta$ values. An $\eta$ value of 0.8 was chosen since this was near the mean of the $\eta$ values found for the freely varying $\eta$ thermal fits. These cases are listed as "fixed W3, W4". Instances of "fixed

W3" or "fixed W4" are cases where only a single thermal band was utilized in the fit. Note that the detection for 2010 $BL_4$ was marginal (3.9 $\sigma$) and only in W4. Also, note that the diameter fit for (10199) Chariklo was smaller than in S08 even though the signal was strong in W3 and W4. However, if a fixed $\eta$ value of ~1.2 is used, similar to that found by S08, the derived diameter, 273 +/- 100 km, and albedo, 0.04 +/- 0.04, values are consistent with the S08 values of 257 +/- 13 km and 0.06 +/-0.01. The diameters for (127546) 2002 $XU_{93}$, 192 +/- 50 km, and (42355) Typhon, 170 +/- 50 km, were consistent with those found in S12, when utilizing the same values of $\eta$, since both bodies were detected only in W4. Finally, it is likely that Chiron was weakly active during its observation, and the implications for Chiron's activity will be discussed in Section 3.

## 3. Discussion

### *3.1 Albedos, Diameters, and Orbital Elements*

We searched our albedo and diameter data for relations based on dynamical properties, with the major caveat that we have not accounted for sample biases that derive from the means by which the objects were detected either by WMOPS or by ground-based visible light surveys. These survey biases can have strong effects on the relationship between these quantities in the observed sample, and caution must be used when interpreting the results without accounting for them. We find no significant relation between size, eccentricity, albedo, or semi-major axis. As stated before, Figure 1 shows only the obvious trends imposed by our detection thresholds. Figure 2 seems at first to indicate that the brightest objects are concentrated at lower inclinations. However, this could as easily be the result of low-number statistics for high-inclination objects, because the ground-based visible surveys that discovered more than half of the objects in our sample do not survey at high inclinations, and so were less likely to find them. Also, high inclination populations may be comprised of objects with cometary origins and surfaces, which have been found to be dark (Fernández et al. 2013). While NEOWISE has been shown to be equally sensitive to low and high albedo objects (Mainzer et al. 2011b, Grav et al. 2011, 2012a) and sensitive to high inclination objects (by virtue of being an all-sky survey), about a third of our sample consists of objects detected by stacking on predicted positions of objects discovered by ground-based visible surveys. These surveys are considerably less likely to discover low albedo objects. The dynamical properties investigated also include the Tisserand-parameter relative to Jupiter ($T_J$), or Neptune ($T_N$), which is defined as:

$$T_{J,N} = \frac{a_{J,N}}{a} + 2\sqrt{\frac{a}{a_{J,N}}(1-e^2)}\,\cos i \qquad (2)$$

where $a_{J,N}$ is the semi-major axis of Jupiter or Neptune, and $a$, $e$, and $i$ are the semi-major axis, eccentricity and inclination of the object. Objects with $T_J < 3$ are considered good candidates for bodies with cometary origins, since most JFCs have $2 < T_J < 3$, and longer period comets have $T_J < 2$ (Levison 1996). Figure 3 is consistent with both the presence of bodies of cometary origin and a possibly significant selection bias, as the low $T_J$ region of the figure tends to be devoid of the higher-albedo objects. These figures indicate that, because the survey biases of the infrared-selected WMOPS-detected sample and the optically-selected stacked sample are quite different, care must be taken when extrapolating the results from this observed sample to the populations of Centaurs, SDOs, and Satyrs as a whole. The process of accounting for these survey biases will be the subject of future work. The collective albedos for the Centaurs yield a mean value of 0.08 +/- 0.04, which is identical for the SDOs as well. The mean albedo of the Satyr sub-population of Centaurs is slightly darker, 0.06 +/- 0.03, but the difference is not statistically significant, and falls within the collective mean of the combined SDO and Centaur sample of 0.08 +/- 0.04 (Figure 4). These values for $p_V$ overlap the average $p_V$ ~0.07 value reported for the samples of Centaurs and SDOs in S08 and S12.

*3.2 Size-Frequency Distributions*

The biased diameters present sufficient first-order statistics for derivation of cumulative size frequency distributions (SFDs; cf. Colwell 1993) for the Centaurs, SDOs and combined samples. Taking the results for all objects, regardless of their distance, we find shallow distributions for the number of these bodies with diameter d > D,

$$N(d > D) \sim D^{-\alpha}$$

with power law indices $\alpha \sim 1$, i.e. SFDs dominated by the sensitivity bias (Figure 5). If, however, we select objects detected between 5 and 10 AU, where the sensitivity gradient is less extreme (and dominated by the W4 response alone), and with diameters greater than 20km, we find SFDs of $\alpha$ = -1.7 +/- 0.2 for the Centaur and $\alpha$ = -1.6 +/- 0.3 for the SDO samples. We also find $\alpha$ = -1.8 +/- 0.2 for the combined sample. These values roughly agree with earlier estimates for the Centaurs ($\alpha$ ~ 2.0; Jedicke et al. 2002). These power law values also compare favorably with the SDF for JFCs of -1.92 +/- 0.2 found by Snodgrass et al. (2011). However, these index values are difficult to interpret without proper de-biasing, which may greatly affect the values. For example, objects at similar distances discovered at visual magnitudes may have higher albedo than those discovered in the thermal IR. In turn, these bodies may have smaller diameters on average, weighting the SFDs towards the smaller-size end and steepening the slope.

### *3.3 Colors and Albedos*

We have found evidence for a correlation between color and surface reflectance in the observed sample. Previous work (cf. Peixinho et al. 2012; Tegler et al. 2008, and references therein) has shown the Centaurs to be divided into two color sub-classes. Most objects tend to have B-R colors closer to the solar value of 1.19 magnitudes (Livingston 2000), i.e. near-neutral reflectance, with B-R colors < 1.4 magnitudes. A smaller group has redder colors with B-R in excess of 1.4 mag, clustering with values near ~1.85 mag. With the limited number of centaur albedos available from S08, Tegler et al. (2008) suggested a possible correlation between B-R colors and surface albedo at the 2-$\sigma$ level based on 15 objects.

**We conclude that this correlation exists using a sample of 41 objects with color and albedo information available, including the eight additional objects from S08 with measured colors, but not observed by WISE (Figure 6).** As a check, we divided our sample into B-R < 1.4 mag ("near-neutral" reflectance, or "bluer", group) and B-R >1.4 mag ("red" group; cf. Tegler et al. 2008), and ran a two-set Kolmogorov-Smirnov test (Press et al. 1992) to determine the probability that the bluer and red surface sub-populations represented different populations with respect to albedo also. With a 99.95% confidence, the two sub-classes were found to be different. The bluer color group had a

mean albedo of 6% +/- 2%, while that of the redder color group was 12% +/- 5%. On average, the redder group had a higher albedo, and also a wider distribution in color. It is also worth noting that the bluer group is closer both in color and albedo to those of cometary nuclei (Fernández et al. 2002; Fernández et al. 2013), so that it seems plausible that a direct link between comets and the bluer subclass of Centaurs exists.

### *3.4 Beaming Parameters*

Further evidence of the link between comets and Centaurs may be indicated by the distribution of beaming parameters (Figure 7). The beaming parameter, or $\eta$, is a value that accounts for deviations of the temperature distribution from instantaneous thermal equilibrium (c.f. Harris & Lagerros 2002, Harris 1998). A body with a low thermal inertia has an $\eta$ value close to 1. Since all objects had $H_V$ estimates (Section 2.1), the beaming parameter was fit for all objects with significant W3 and W4 signal. Those with successful beaming parameter fits are labeled as "Free" in the fits comments column in Table 4. The distribution of the $\eta$ values from fits of beaming parameter values for objects with significant signal at two or more wavelengths is plotted in Figure 7. The mean value for the total sample and the Centaurs, $\eta$ = 0.9 +/- 0.2, is identical to those found for comets (Fernández et al. 2013; Bauer et al. 2012c), including the dispersion as reflected by the standard deviation. For the SDOs, the mean value of $\eta$ = 0.83 +/- 0.10 lies within the dispersion of the Centaur average beaming parameters. Similarly, the red ($\eta$ = 0.73 +/- 0.11) and bluer ($\eta$ = 0.87 +/- 0.18) Centaur beaming parameters overlap statistically.

*3.5 Chiron and the Active Centaurs*

Four Centaurs with confirmed observations of activity were observed by the WISE spacecraft: 174P/Echeclus, C/2010 $KW_7$, 95P/Chiron, and 29P/Schwassmann-Wachmann 1. Comet 29P/SW1 was active at the time, and a detailed analysis of the 29P observations is the subject of another paper. We present here a diameter value of 46 +/- 13 km for 29P's nucleus from a preliminary analysis (Fernández et al. 2012) using the extraction technique described in Bauer et al. (2012b), which is consistent with the value of 54+/- 10 km from Stansberry et al. (2004). Our fitted value for Centaur 174P/Echeclus has a diameter of 59 +/- 4 km within 1.7σ of that derived in S08, 84 +/- 15 km. We derived an albedo for 174P of 0.08 +/- 0.02, nearly twice that found by S08 (0.04 +/- 0.02). Activity in 174P was previously detected at wavelengths of 24 and 70 μm using SST (Bauer et al. 2008). There was no indication of activity for 174P in the WISE images or reported in S08, though it cannot be completely ruled out. C/2010 KW7, with a diameter of 6 +/- 1 km based on the NEATM thermal fit, did not show any signs of activity until after its discovery by NEOWISE (cf. Scotti & Williams 2010).

Chiron may have been active at the time of the WISE observations, but the WISE data are consistent with no or low activity. The diameter derived from the adapted NEATM thermal fits for the Chiron data (201 +/- 62 km) is in line with other nucleus size estimates such as in S08 (233 +/- 15 km) and including that from a stellar occultation (180 +/- 14 ; Bus et al. 1996). However, magnitudes at visual wavelengths reported at or around the time of the NEOWISE observations at the MPC, and our own observations, were brighter than those expected for a bare nucleus. We observed Chiron on 22 May 2010 from the Palomar Transient Factory's 48-inch telescope (PTF; cf. Law et al. 2008;

Waszczak et al 2013; Ofek et al. 2012a,b) and found an R-band magnitude ($m_R$) = 17.86 +/- 0.03. For a bare nucleus of ~200km diameter with ~15% reflectance (Fernández et al. 2002) and an IAU phase curve slope parameter of G=0.15 (cf. Bowell et al. 1989), we derive $m_R$ ~18.2. Our measured magnitude is thus consistent with excess brightness caused by activity at the ~0.3 mag level. Comet dust is usually red (Jewitt & Meech, 1986), but in the case of Chiron, the coma is neutral (Bauer et al. 1997), which can be explained by the presence of small grains (Meech et al. 1997). If the excess brightness in the PTF observations was caused by mostly small-grained dust (cf. Meech et al. 1997, Bauer et al. 1997), the excess would not be readily apparent at the W3 and W4 wavelengths, as shown below.

We provide a fit to the flux data using the method applied to 67P/Churymov-Gerasimenko (Bauer et al. 2012b), with the visual-band constraint provided by the PTF data, shown in Figure 8. We assume an $\alpha$ = 3 dust particle size frequency distribution (DPSFD; cf. Fulle et al. 2004), with both a neutral-reflectance and a reflectance reddening law based on Jewitt & Meech (1986) averaged out to 3.5μm. A value of $\alpha$ near 3 would be consistent with common cometary DPSFDs, but may overestimate the dust flux at W1 and W2 for Chiron, possibly owing to an over-abundance of smaller grains. The case for the presence of small grains is possibly strengthened by the temperature fits of 101 +/- 5K, well in excess of the 69K black-body temperature, but nearly matching the sub-solar-point temperature of ~98K for $R_h$ ~ 16.3 AU. Higher temperature fits to dust coma in excess of black body and sub-solar-point temperatures have been found to be consistent with the dominance of small dust grains in the coma (cf. Lisse et al. 1998;

Reach et al. 2000), owing to super-heating and the presence of silicate emission at ~10-12 μm. We find a weak W2 signal that is in slight excess of the dust signal that can accommodate CO or $CO_2$ production rates of $1 \times 10^{28}$ and $1 \times 10^{27}$ mol/sec, respectively (1σ upper limits; Bauer et al. 2012b, Pittichova et al. 2008). However, the excess 4.6 μm flux is greater than the predicted dust and measured nucleus signal at barely the 1 σ level (Figure 8), and may as likely be caused by dust scattering or other effects. It is therefore not considered a reliable detection of either species.

## 4. Conclusions

We find 52 Centaur and SDO objects in the WISE data. Thermal fits yield diameters and albedos for these objects reported in Table 4. We combine known diameters, albedos, and colors from Stansberry et al. 2008 and Tegler et al. 2008 with the WISE data and our own color photometry obtained from ground-based observations (Table 2), and find that red and blue Centaurs reflect different populations with respect to albedo with a 99.95% confidence. Blue objects have mean albedos of 0.06 +/- 0.02 and redder objects have mean albedos of 0.12 +/- 0.05. NEATM thermal fits yield mean beaming parameters over the entire sample of 0.89 +/- 0.17, similar to those reported for comet populations, and are consistent with low thermal-inertia surfaces. Cumulative SFDs indicate power law slope values agree similar to earlier estimates for the Centaurs and for the JFCs, with α ~ −1.7 +/− 0.2. However, this sample is not debiased, so the SFDs are probably not representative of the general population, given the various selection effects at work.

## 5. Acknowledgements

This publication makes use of data products from the Wide-field Infrared Survey Explore, which is a joint project of the University of California, Los Angeles, and the Jet Propulsion Laboratory/California Institute of Technology, funded by the National Aeronautics and Space Administration. This publication also makes use of data products from NEOWISE, which is a project of JPL/Caltech, funded by the Planetary Science Division of NASA. This material is based in part upon work supported by the NASA through the NASA Astrobiology Institute under Cooperative Agreement No. NNA09DA77A issued through the Office if Space Science. R. Stevenson is supported by the NASA Postdoctoral Program, and E. Kramer acknowledges her support through the JPL graduate internship program. Data were based in part on observations obtained at the Hale Telescope, Palomar Observatory, as part of a collaborative agreement between the California Institute of Technology, its divisions Caltech Optical Observatories and the Jet Propulsion Laboratory (operated for NASA). Also, this work is based in part on observations obtained at the Southern Astrophysical Research (SOAR) telescope, which is a joint project of the Ministério da Ciência, Tecnologia, e Inovação (MCTI) da República Federativa do Brasil, the U.S. National Optical Astronomy Observatory (NOAO), the University of North Carolina at Chapel Hill (UNC), and Michigan State University (MSU), with time allocated through NOAO. We thank the anonymous reviewer for the very helpful comments of manuscript drafts.

**Table 1: Extracted Fluxes of Centaurs and SDOs**

| Object[a] | MPC Designation[1] | $H_V$[a] | $R_h$ [AU][a] | Δ [AU][a] | α[°][a] | W3 [mJy] | W4 [mJy] | W1 [mJy] | W2 [mJy] |
|---|---|---|---|---|---|---|---|---|---|
| 10199 | (10199) Chariklo | 6.66 | 13.75 | 13.72 | 4.1 | 0.5±.1 | 52±5 | -- | -- |
| 167P | 167P/CINEOS | 9.7 | 14.31 | 14.28 | 4.1 | -- | 5±1 | -- | -- |
| 2060 | (2060) Chiron | 6.5 | 16.31 | 16.28 | 3.6 | 0.30±.06 | 24±3 | -- | 0.08±.01 |
| 29P [b] | 29P/Schwassmann-Wachmann 1 | 9. | 6.21 | 6.04 | 9.3 | 18±3 | 144±23 | -- | -- |
| 31824 | (31824) Elatus | 10.1 | 14.28 | 14.15 | 4.0 | -- | 4±1 | -- | -- |
| 32532 | (32532) Thereus | 9. | 11.8 | 11.67 | 4.8 | 0.19±.03 | 6.0±.1 | .11±.03 | .14±.03 |
| 42355 | (42355) Typhon | 7.5 | 18.08 | 17.96 | 3.2 | -- | 4±1 | -- | -- |
| 52872 | (52872) Okyrhoe | 11. | 6.72 | 6.54 | 8.7 | 2.9±.3 | 42±4 | -- | -- |
| 52872 | (52872) Okyrhoe | 11. | 6.44 | 6.35 | 8.8 | 3.5±.4 | 47±4 | -- | -- |
| 54598 | (54598) Bienor | 7.5 | 17.37 | 17.33 | 3.4 | 0.05±.01 | 11±1 | -- | -- |
| 55576 | (55576) Amycus | 7.46 | 16.72 | 16.69 | 3.4 | -- | 5±1 | -- | -- |
| 60558 | (60558) Echeclus | 9.55 | 9.77 | 9.71 | 5.8 | 1.5±.3 | 32±5 | -- | -- |
| 60558 | (60558) Echeclus | 9.55 | 9.32 | 9.17 | 6.2 | 1.8±.3 | 37±5 | -- | -- |
| 95626 | (95626) 2002 GZ32 | 7. | 19.12 | 19.11 | 2.9 | -- | 15±2 | -- | -- |
| 95626 | (95626) 2002 GZ32 | 7. | 19.02 | 18.92 | 3.0 | -- | 15±1 | -- | -- |
| C0061 | (120061) 2003 CO1 | 8.9 | 11.80 | 11.76 | 4.8 | 0.4±.1 | 19±2 | -- | -- |
| C7546 | (127546) 2002 XU93 | 7.9 | 21.06 | 20.95 | 2.7 | -- | 2.5±.4 | -- | -- |
| CK11K36P | C/2011 KP36 (Spacewatch) | 9.4 | 14.54 | 14.51 | 3.9 | -- | 3.3±.7 | -- | -- |
| D6204 | (136204) 2003 WL7 | 8.6 | 14.98 | 14.85 | 3.8 | -- | 9±2 | -- | -- |
| E5486 | (145486) 2005 UJ438 | 10.5 | 8.27 | 8.14 | 6.9 | 0.5±.1 | 4±1 | -- | -- |
| E8975 | (148975) 2001 XA255 | 11.2 | 9.37 | 9.25 | 6.0 | -- | 11±4 | -- | -- |
| E8975 | (148975) 2001 XA255 | 11.2 | 9.34 | 9.21 | 6.2 | 0.8±.2 | 15±3 | -- | -- |
| K05VB9J | 2005 VJ119 | 10.6 | 11.51 | 11.37 | 5.0 | -- | 2.5±.5 | -- | -- |
| K05VB9J | 2005 VJ119 | 10.6 | 11.38 | 11.31 | 5.0 | -- | 3.2±.8 | -- | -- |
| K07VU5H | 2007 VH305 | 8.21 | 8.21 | 8.09 | 6.9 | 1.0±.2 | 6±1 | -- | -- |
| K08H21Y | 2008 HY21 | 12.1 | 6.94 | 6.86 | 8.3 | 0.6±.1 | 12±2 | -- | -- |
| K08J14S | 2008 JS14 | 13.2 | 5.45 | 5.28 | 10.4 | 1.4±.4 | 14±3 | -- | -- |
| K10B04L | 2010 BL4 | 11.9 | 8.65 | 8.49 | 6.6 | -- | 3.1±.8 | -- | -- |
| K10BB8K | 2010 BK118 | 10.2 | 8.22 | 8.09 | 6.7 | 1.9±.2 | 33±5 | -- | -- |
| K10CE0R | 2010 CR140 | 15.5 | 4.20 | 4.04 | 13.6 | 1.6±.1 | 10±1 | -- | -- |
| K10E65S | 2010 ES65 | 11.8 | 9.69 | 9.56 | 6.0 | 0.30±.07 | 5.4±.9 | -- | -- |
| K10F92H | 2010 FH92 | 11.7 | 5.79 | 5.70 | 9.9 | 7.5±.5 | 60±4 | -- | -- |
| K10G64W | 2010 GW64 | 14.9 | 3.71 | 3.50 | 15.6 | 4.2±.4 | 16±2 | -- | -- |
| K10GE7W | 2010 GW147 | 13.2 | 6.09 | 6.01 | 9.5 | 1.2±.1 | 12.6±0.9 | -- | -- |
| K10H20U | 2010 HU20 | 13.0 | 4.74 | 4.63 | 12.2 | 1.9±.3 | 13±2 | -- | -- |
| K10JC4H | 2010 JH124 | 14.6 | 4.13 | 3.93 | 14.1 | 2.1±.3 | 11±2 | -- | -- |
| K10K07W | 2010 KW7 | 15.5 | 3.00 | 2.74 | 19.6 | 5.7±.5 | 17±2 | -- | -- |
| K10L61G | 2010 LG61 | 18.5 | 1.57 | 1.21 | 40.3 | 3.6±.5 | 8±3 | -- | -- |
| K10O01R | 2010 OR1 | 16.2 | 2.78 | 2.54 | 20.7 | 4.2±.5 | 12±2 | 0.41±.03 | -- |
| K10OA1M | 2010 OM101 | 17. | 2.25 | 2.01 | 26.8 | 7.7±.8 | 19±3 | -- | -- |
| K10P58O | 2010 PO58 | 14.5 | 4.50 | 4.33 | 13.0 | 2.1±.2 | 13±2 | -- | -- |
| K10R64M | 2010 RM64 | 11.0 | 7.79 | 7.71 | 7.5 | 0.46±.09 | 8±2 | -- | -- |
| K10T00H | 2010 TH | 9.2 | 13.99 | 13.96 | 4.2 | -- | 7±1 | -- | -- |
| K10W09G | 2010 WG9 | 8.1 | 19.33 | 19.24 | 2.9 | -- | 2.9±.3 | -- | -- |
| K11M04M | 2011 MM4 | 9.3 | 12.94 | 12.90 | 4.4 | 0.11±.02 | 7.3±.7 | -- | -- |
| O8835 | (248835) 2006 SX368 | 9.5 | 11.96 | 11.91 | 4.9 | 0.6±.2 | 18±2 | -- | -- |
| P0112 | (250112) 2002 KY14 | 9.5 | 8.73 | 8.69 | 6.7 | 1.2±.2 | 22±3 | -- | -- |
| S1371 | (281371) 2008 FC76 | 9.1 | 10.95 | 10.89 | 5.3 | 0.8±.1 | 21±2 | -- | -- |
| U9139 | (309139) 2006 XQ51 | 9.8 | 12.85 | 12.73 | 4.4 | -- | 3.0±.7 | -- | -- |
| U9737 | (309737) 2008 SJ236 | 12.2 | 6.60 | 6.44 | 8.6 | 1.1±.2 | 13±2 | -- | -- |
| V0071 | (310071) 2010 KR59 | 7.7 | 13.00 | 12.85 | 4.4 | -- | 22±3 | -- | -- |
| W8884 | (328884) 2010 LJ109 | 10.1 | 9.14 | 8.99 | 6.4 | 1.0±.2 | 21±3 | -- | -- |
| X0759 | (330759) 2008 SO218 | 12.8 | 3.70 | 3.57 | 15.9 | 12±1 | 52±5 | -- | -- |
| X2685 | (332685) 2009 HH36 | 10.6 | 8.02 | 7.71 | 6.8 | 1.3±.2 | 22±3 | -- | -- |
| X6756 | (336756) 2010 NV1 | 10.5 | 9.45 | 9.42 | 6.2 | 1.0±.3 | 21±5 | -- | -- |
| Y2842 | (342842) 2008 YB3 | 9.5 | 6.65 | 6.51 | 8.7 | 14±1 | 171±10 | 0.16±.02 | -- |
| Y6889 | (346889) 2009 QV38 | 11.8 | 6.05 | 5.95 | 9.6 | 3.2±.5 | -- | -- | -- |

[a]Object name in packed-provisional format (http://www.minorplanetcenter.org/iau/info/PackedDes.html). The MPC Designation is the full designation as

specified in the Minor Planet Center's one-line format ephemeris output for the object. $H_V$ – Solar-system absolute magnitude; Rh – heliocentric distance of object at time of observation in AU; Δ – object-spacecraft distance at time of observation in AU. α – Sun-object-spacecraft (phase) angle in degree units. [b] Preliminary coma-extracted fluxes for 29P are provided here (see text).

**Table 2: New Centaur & SDO Colors**

| Object | Date (UT) | Telescope / Instrument | $m_R$ | B - R |
|---|---|---|---|---|
| C2011K36P | 3/27, 2012 | Pal. 200in/LFC | 20.88 ± .08 | 1.42 ± .10 |
| E5486 | 3/21/2012 | SOAR/SOI | 20.92 ± .04 | 1.55 ± .05 |
| K07VU5H | 9/19, 2012 | Pal. 200in/LFC | 20.14 ± .01 | 2.27 ± .09 |
| K10 BB8K | 7/10, 2012 | Pal. 200in/LFC | 18.27 ± .02 | 1.29 ± .02 |
| K10B41L | 3/21, 2012 | SOAR/SOI | 21.57 ± .11 | 1.09 ± .14 |
| K10F92H | 4/18, 2010 | SOAR/SOI | 19.22 ± .03 | 1.30 ± .05 |
| K10GE47W | 3/27, 2012 | Pal. 200in/LFC | 23.20 ± .10 | 0.95 ± .14 |
| K10R64M | 3/21, 2012 | SOAR/SOI | 18.52 ± .02 | 2.24 ± .05 |
| O8835 | 9/19, 2012 | Pal. 200in/LFC | 20.17 ± .03 | 1.28 ± .14 |
| P0112 | 9/19, 2012 | Pal. 200in/LFC | 19.34 ± .03 | 1.69 ± .04 |
| S1371 | 9/19, 2012 | Pal. 200in/LFC | 19.18 ± .03 | 1.52 ± .07 |
| V0071 | 3/21, 2012 | SOAR/SOI | 20.91 ± .12 | 1.59 ± .12 |
| V8884 | 5/26, 2012 | Pal. 200in/LFC | 19.73 ± .03 | 1.19 ± .17 |
| X0759 | 3/21, 2012 | SOAR/SOI | 20.98 ± .04 | 1.50 ± .10 |
| X2685 | 3/21, 2012 | SOAR/SOI | 19.84 ± .03 | 1.10 ± .04 |
| X6756 | 7/10, 2012 | Pal. 200in/LFC | 20.42 ± .10 | 1.20 ± .14 |
| Y6889 | 11/12, 2012 | Pal. 200in/LFC | 20.75 ± .04 | 1.15 ± .14 |

**Table 3: Known Centaur Colors**

| **Object** | **Reference** | **B - R** |
|---|---|---|
| 10199 | Tegler et al. 2008 | 1.26 ± .04 |
| 10370 | Tegler et al. 2008 | 1.15 ± .06 |
| 167P | Jewitt 2009 | 1.30 ± .05 |
| 2060 | Tegler et al. 2008 | 1.04 ± .05 |
| 29P | Jewitt 2009 | 1.28 ± .04 |
| 31824 | Tegler et al. 2008 | 1.70 ± .02 |
| 32532 | Tegler et al. 2008 | 1.18 ± .01 |
| 42355 | Peixinho et al. 2012 | 1.29 ± .07 |
| 5145 | Tegler et al. 2008 | 2.04 ± .07 |
| 52872 | Tegler et al. 2008 | 1.21 ± .02 |
| 52975 | Tegler et al. 2008 | 1.86 ± .05 |
| 54598 | Tegler et al. 2008 | 1.12 ± .03 |
| 55576 | Tegler et al. 2008 | 1.79 ± .03 |
| 60558 | Tegler et al. 2008 | 1.38 ± .04 |
| 63252 | Tegler et al. 2008 | 1.20 ± .03 |
| 7066 | Tegler et al. 2008 | 1.88 ± .06 |
| 83982 | Tegler et al. 2008 | 1.85 ± .02 |
| 8405 | Tegler et al. 2008 | 1.23 ± .05 |
| 95626 | Tegler et al. 2008 | 1.03 ± .04 |
| C0061 | Tegler et al. 2008 | 1.24 ± .04 |
| C7546 | Hainaut et al. 2012 | 1.20 ± .03 |
| D6204 | Peixinho et al. 2012 | 1.23 ± .04 |
| E8975 | Fraser & Brown 2012 | 1.23 ± .10[a] |
| Y2842 | Sheppard 2010 | 1.26 ± .01 |

[a]B-R color extrapolated from Hubble Space Telescope's Wide-Field Camera 3 observations through the F606w and F814w filters.

**Table 4: NEATM Model Fits to WISE Flux Measurements.**

| Objecta | Diam | σ - Diam | Albedo | σ - Albedo | η | σ - η | NEATM fit Comments | e | a | i | q | $T_N$ | WMOPS-Observed? |
|---|---|---|---|---|---|---|---|---|---|---|---|---|---|
| **Centaurs** | | | | | | | | | | | | | |
| *"Satyr" Centaurs, 5.5 AU < a < 30.1, q < 5.2* | | | | | | | | | | | | | |
| K08J14S | 14.5 | 1.8 | 0.044 | 0.019 | 1.046 | 0.186 | Free | 0.742 | 11.65 | 26.0 | 3.0 | 3.28 | N |
| K10CE0R | 7.5 | 1.4 | 0.020 | 0.01 | 1.111 | 0.283 | Free | 0.408 | 5.62 | 74.7 | 3.32 | 5.67 | D,Y |
| K10H20U | 10.513 | 1.1 | 0.101 | 0.024 | 0.976 | 0.162 | Free | 0.2719 | 5.8452 | 22.43 | 4.2558 | 4.33 | D,Y |
| K10L61G | 0.89 | 0.19 | 0.089 | 0.056 | 1.00 | 0.400 | Free | 0.8078 | 7.1115 | 123.7 | 1.367 | 3.97 | D,Y |
| K10O01R | 3.25 | 0.64 | 0.055 | 0.013 | 0.831 | 0.146 | Free | 0.9245 | 27.188 | 143.8 | 2.0516 | 1.13 | D,Y |
| K10OA1M | 3.12 | 0.17 | 0.029 | 0.005 | 1.054 | 0.105 | Free | 0.9193 | 26.401 | 118.7 | 2.129 | 1.12 | D,Y |
| K10P58O | 8.88 | 0.63 | 0.035 | 0.007 | 0.915 | 0.093 | Free | 0.6594 | 8.8008 | 121.1 | 2.9979 | 3.28 | D,Y |
| X0759 | 11.8 | 0.4 | 0.097 | 0.017 | 0.823 | 0.046 | Free | 0.563 | 8.12 | 170.4 | 3.5 | 4.29 | N |
| *Centaurs, 5.5 AU < a < 30.1, q > 5.2* | | | | | | | | | | | | | |
| 10199 | 226.1 | 29.3 | 0.075 | 0.015 | 1.009 | 0.049 | Free | 0.170 | 15.73 | 23.4 | 13.1 | 0.86 | N |
| 167P | 66.17 | 22.9 | 0.053 | 0.019 | 0.8 | 0.360 | Fixed η W4 | 0.2700 | 16.141 | 19.13 | 11.784 | 0.54 | N |
| 2060 | 201.2 | 62.4 | 0.110 | 0.052 | 0.8 | 0.320 | Fixed η W3, W4 | 0.3792 | 13.670 | 6.9 | 8.486 | 3.29 | N |
| 29P | 46. | 13. | 0.033 | 0.015 | 0.64 | 0.29 | Free | 0.044 | 5.998 | 9.38 | 5.723 | 4.15 | N |
| 31824 | 57.0 | 15.9 | 0.050 | 0.028 | 0.8 | 0.272 | Fixed η W4 | 0.3835 | 11.788 | 5.24 | 7.267 | 2.79 | N |
| 32532 | 86.5 | 1.9 | 0.059 | 0.013 | 1.325 | 0.049 | Free | 0.197 | 10.663 | 20.3 | 8.561 | 3.70 | N |
| 52872 | 36.0 | 1.2 | 0.058 | 0.02 | 0.97 | 0.06 | Free | 0.306 | 8.34 | 15.66 | 5.792 | 3.93 | Y |
| 54598 | 187.5 | 15.5 | 0.050 | 0.019 | 1.0 | 0.09 | Free | 0.200 | 16.561 | 20.7 | 13.256 | 1.93 | N |
| 55576 | 100.9 | 40.1 | 0.180 | 0.135 | 0.8 | 0.418 | Fixed η W4 | 0.392 | 24.99 | 13.3 | 15.192 | 0.87 | N |
| 60558 | 59. | 4. | 0.077 | 0.015 | 0.67 | 0.07 | Free | 0.457 | 10.71 | 4.34 | 5.81 | 1.87 | Y |
| 95626 | 230.5 | 87.50 | 0.053 | 0.030 | 0.8 | 0.343 | Fixed η W4 | 0.217 | 23.024 | 15.0 | 17.991 | 1.67 | N |
| C0061 | 82.0 | 8.4 | 0.072 | 0.032 | 0.851 | 0.126 | Free | 0.473 | 20.7 | 19.8 | 10.92 | 0.08 | N |
| D6204 | 118.0 | 32.8 | 0.046 | 0.029 | 1.0 | 0.357 | Fixed η W4 | 0.261 | 20.22 | 11.2 | 14.95 | 2.56 | N |
| E5486 | 20.8 | 7.2 | 0.215 | 0.123 | 0.8 | 0.395 | Fixed η W3, W4 | 0.532 | 17.637 | 3.8 | 8.258 | 0.96 | N |
| E8975 | 37.7 | 10.5 | 0.041 | 0.014 | 0.703 | 0.186 | Free | 0.677 | 28.9 | 12.6 | 9.33 | 2.20 | N |
| K07VU5H | 23.8 | 8.0 | 0.070 | 0.036 | 0.8 | 0.384 | Fixed η W3, W4 | 0.667 | 24.606 | 6.2 | 8.188 | -0.09 | N |
| K08H21Y | 24.0 | 1.5 | 0.044 | 0.010 | 1.22 | 0.094 | Free | 0.507 | 10.96 | 12 | 5.41 | 1.81 | N |
| K10B04L | 15.7 | 3.2 | 0.114 | 0.052 | 0.8 | 0.333 | Fixed η W4 | 0.5393 | 18.612 | 20.81 | 8.573 | 1.74 | N |
| K10E65S | 26.9 | 7.9 | 0.049 | 0.024 | 0.8 | 0.28 | Fixed η W3, W4 | 0.556 | 21.348 | 10.4 | 9.47 | 2.14 | N |
| K10F92H | 28.0 | 0.6 | 0.047 | 0.007 | 0.730 | 0.023 | Free | 0.763 | 24.4 | 61.9 | 5.785 | 0.10 | D,Y |
| K10R64M | 21.0 | 2.0 | 0.159 | 0.048 | 0.85 | 0.144 | Free | 0.685 | 19.57 | 27. | 6.16 | 1.82 | N |
| K10T00H | 69.9 | 24.2 | 0.078 | 0.033 | 0.8 | 0.363 | Fixed η W4 | 0.3254 | 18.659 | 26.69 | 12.587 | 0.50 | N |
| K11M04M | 63.7 | 6.2 | 0.083 | 0.024 | 0.841 | 0.119 | Free | 0.482 | 21.51 | 100.4 | 11.1 | -0.03 | N |
| O8835 | 78.44 | 22.63 | 0.046 | 0.018 | 0.8 | 0.357 | Fixed η W3, W4 | 0.463 | 22.288 | 36.3 | 11.961 | 2.87 | N |
| P0112 | 38.9 | 3.5 | 0.185 | 0.046 | 0.661 | 0.094 | Free | 0.316 | 12.6 | 19.46 | 8.62 | 1.21 | D,Y |
| S1371 | 58.0 | 4.2 | 0.120 | 0.027 | 0.586 | 0.061 | Free | 0.311 | 14.786 | 27.1 | 10.18 | 2.99 | N |
| U9139 | 39.1 | 15.7 | 0.139 | 0.058 | 0.8 | 0.456 | Fixed η W4 | 0.3779 | 15.920 | 31.57 | 9.903 | 2.90 | N |
| U9737 | 17.7 | 1.5 | 0.074 | 0.021 | 0.800 | 0.110 | Free | 0.439 | 10.99 | 6.0 | 6.2 | 2.49 | N |
| V0071 | 110.06 | 30.82 | 0.121 | 0.037 | 0.8 | 0.324 | Fixed η W4 | 0.5658 | 29.902 | 19.67 | 12.984 | -0.12 | N |
| W8884 | 44.2 | 3.8 | 0.083 | 0.021 | 0.748 | 0.103 | Free | 0.324 | 13.5 | 24.8 | 9.13 | 2.98 | D,Y |
| X2685 | 33.0 | 2.8 | 0.078 | 0.018 | 0.739 | 0.095 | Free | 0.446 | 12.7 | 23.3 | 7.0 | 1.23 | N |
| Y2842 | 67.1 | 1.0 | 0.062 | 0.012 | 0.839 | 0.012 | Free | 0.443 | 11.7 | 105.0 | 6.49 | 1.46 | Y |
| Y6889 | 23.2 | 9.5 | 0.062 | 0.049 | 0.8 | 0.389 | Fixed η W3 | 0.4483 | 10.890 | 19.88 | 6.008 | 2.53 | Y |
| **Scattered Disk Objects** | | | | | | | | | | | | | |
| 42355 | 192. | 50. | 0.05 | 0.03 | 1.48 | 0.4 | Fixed η W4 | 0.535 | 37.633 | 2.4 | 17.516 | 2.22 | N |
| C7546 | 170. | 50. | 0.04 | 0.03 | 1.1 | 0.4 | Fixed η W4 | 0.6858 | 66.784 | 77.90 | 20.984 | -0.99 | N |
| CK11K36P | 55.1 | 19.4 | 0.101 | 0.062 | 0.8 | 0.4 | Fixed η W4 | 0.875 | 38.93 | 18.98 | 4.88 | 1.75 | N |
| K05VB9J | 28.5 | 6.9 | 0.126 | 0.060 | 0.8 | 0.30 | Fixed η W4 | 0.6791 | 35.104 | 6.954 | 11.264 | -0.50 | N |
| K10BB8K | 46.4 | 1.8 | 0.068 | 0.013 | 0.821 | 0.043 | Free | 0.986 | 446.8 | 143.9 | 6.105 | 0.34 | D,Y |
| K10G64W | 6.42 | 0.38 | 0.047 | 0.012 | 0.795 | 0.075 | Free | 0.9416 | 63.459 | 105.3 | 3.7078 | 0.93 | D,Y |
| K10GE7W | 15.9 | 0.7 | 0.037 | 0.006 | 0.869 | 0.056 | Free | 0.973 | 199.3 | 99.7 | 5.38 | 0.80 | D,Y |
| K10JC4H | 7.04 | 0.74 | 0.052 | 0.024 | 0.959 | 0.164 | Free | 0.9694 | 85.344 | 53.37 | 2.6132 | 0.18 | D,Y |
| K10K07W | 4.87 | 0.22 | 0.047 | 0.011 | 0.75 | 0.06 | Free | 0.9684 | 81.000 | 147.1 | 2.5615 | 0.62 | D,Y |
| K10W09G | 112.7 | 61.9 | 0.074 | 0.080 | 0.8 | 0.423 | Fixed η W4 | 0.6511 | 53.747 | 70.21 | 18.753 | 0.68 | N |
| X6756 | 44.2 | 8.0 | 0.057 | 0.030 | 0.661 | 0.168 | Free | 0.968 | 294.0 | 140.8 | 9.417 | 1.58 | D,Y |
| ***Supplemental Data*** | | | | | | | | | | | | | |
| *NEATM fits from Stansberry et al. 2008* | | | | | | | | | | | | | |
| *5145* | *98.* | *25.* | *0.16* | *0.06* | *1.2* | *0.35* | *S08 Fit, 24μm* | *0.571* | *20.25* | *24.71* | *8.69* | *0.84* | *X* |
| *7066* | *60.* | *15.* | *0.06* | *0.04* | *1.2* | *0.35* | *S08 Fit, 24μm* | *0.524* | *24.83* | *15.63* | *11.81* | *-0.31* | *X* |
| *8405* | *85.* | *12.* | *0.05* | *0.02* | *0.7* | *0.2* | *S08 Fit, 2-band*[b] | *0.622* | *18.16* | *17.61* | *6.86* | *0.61* | *X* |
| *10370* | *70.5* | *19.* | *0.06* | *0.04* | *1.2* | *0.35* | *S08 Fit, 24μm* | *0.249* | *25.11* | *4.14* | *18.86* | *1.23* | *X* |
| *31824* | *41.* | *8.* | *0.06* | *0.04* | *1.2* | *0.35* | *S08 Fit, 24μm* | *0.414* | *12.74* | *5.59* | *7.46* | *3.53* | *X* |
| *52975* | *62.* | *16.* | *0.12* | *0.06* | *1.2* | *0.35* | *S08 Fit, 24μm* | *0.384* | *26.41* | *12.62* | *16.26* | *2.65* | *X* |
| *63252* | *34.* | *7.* | *0.04* | *0.02* | *1.2* | *0.35* | *S08 Fit, 24μm*[b] | *0.298* | *9.79* | *12.47* | *6.87* | *2.83* | *X* |
| *83982* | *59.* | *13.* | *0.11* | *0.07* | *1.2* | *0.35* | *S08 Fit, 24μm* | *0.274* | *19.34* | *12.78* | *14.04* | *0.06* | *X* |

aObject name in packed-provisional format (http://www.minorplanetcenter.org/iau/info/PackedDes.html). $T_N$ refers to the Tisserand Parameter for Neptune.

[b] Listed diameters, albedos and $\eta$, along with their uncertainties, are from S08. 2-band fits were from flux at 24 and 70μm, 24μm fits used fixed $\eta$=1.2+.-0.35.

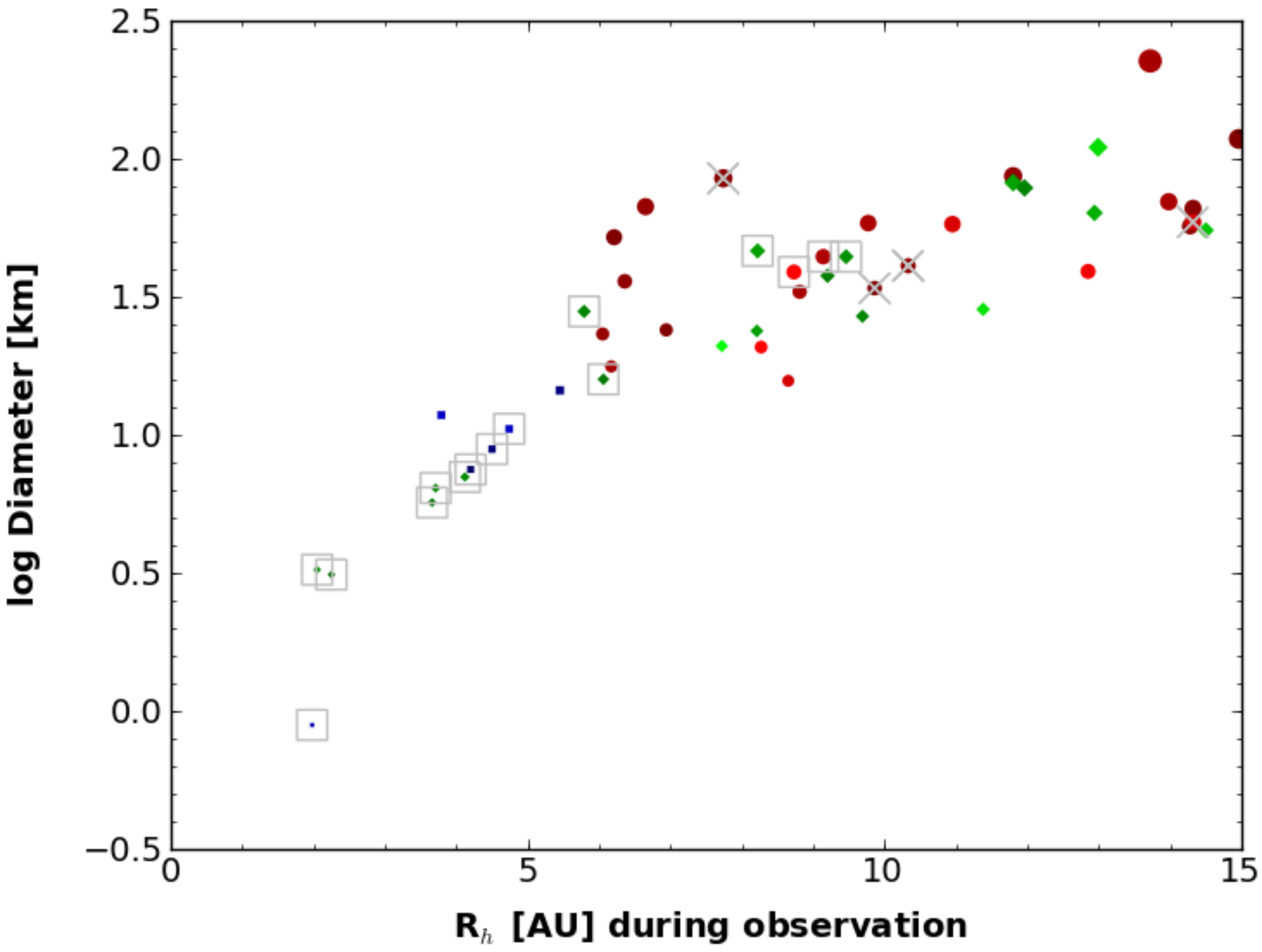

Figure 1: The log of the fitted diameter as compared to the heliocentric distance at the time of observations for 44 of the objects in our survey, plus four from S08 and Tegler et al. 2008. Symbols are color-coded as SDOs (green diamonds), Centaurs (red circles), and the Centaur sub-population of "Satyrs" (blue squares) with q<5.2. The log value of the fitted diameter is also represented by the size of the symbol. The relative visual-band albedo is shown approximately as the darkness of the symbol. Those symbols with boxes around them indicate the objects discovered by NEOWISE, and those with an "X" through them represent the supplemented known objects not detected by WISE, but with measured diameters and colors reported in Stanberry et al. 2008 and Tegler et al. 2008.

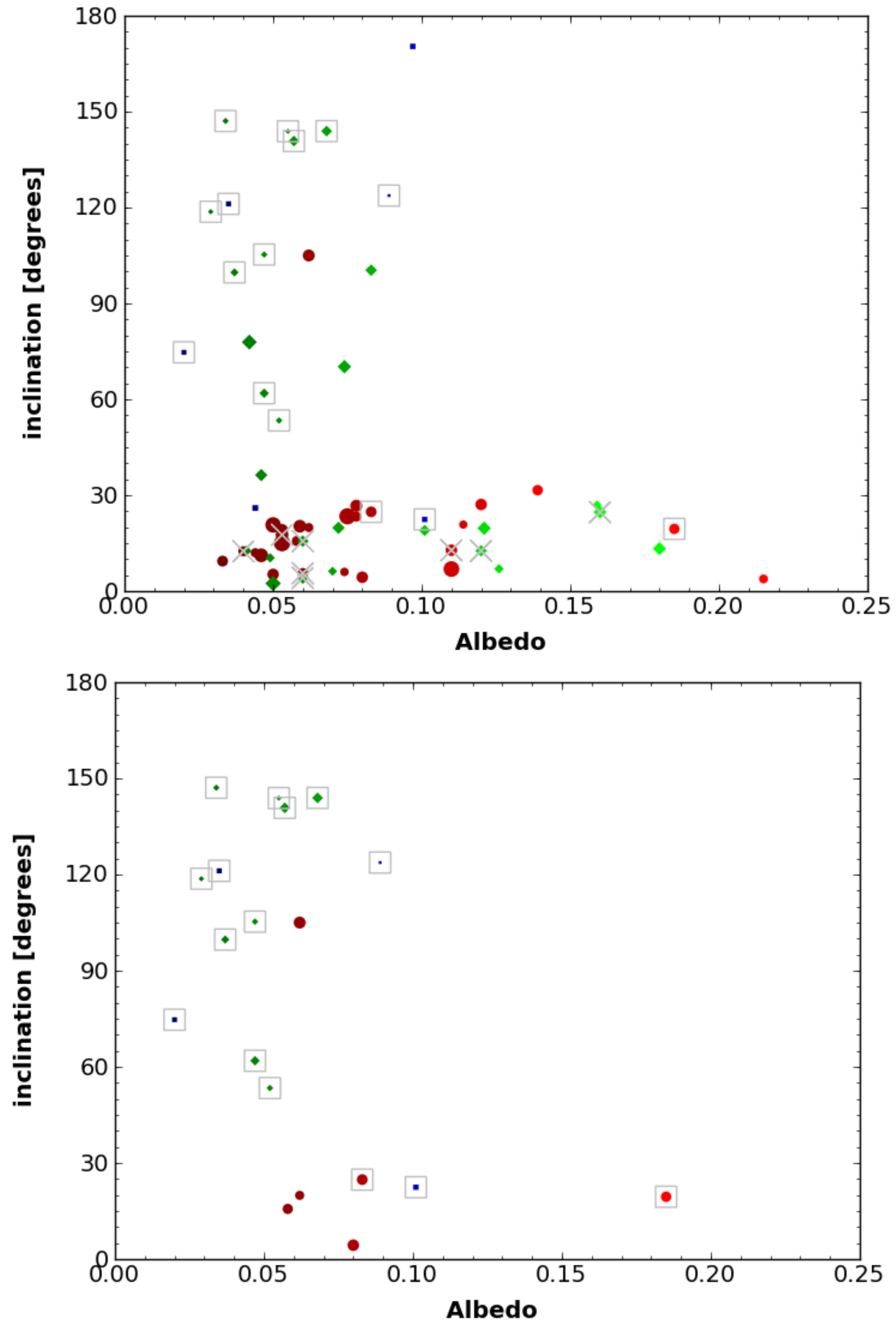

Figure 2: The albedo and orbital inclination, *i*. The symbols are as in Figure 1, with the log value of the fitted diameter represented by the size of the symbol, and the relative visual-band albedo shown approximately as the darkness of each symbol. The inclination of the populations are shown for the entire sample (top) and the WMOPS-detected sub-sample (bottom). Comparison between the top and bottom panels demonstrates the possible selection bias with respect to inclination that may be extant in ground-based surveys.

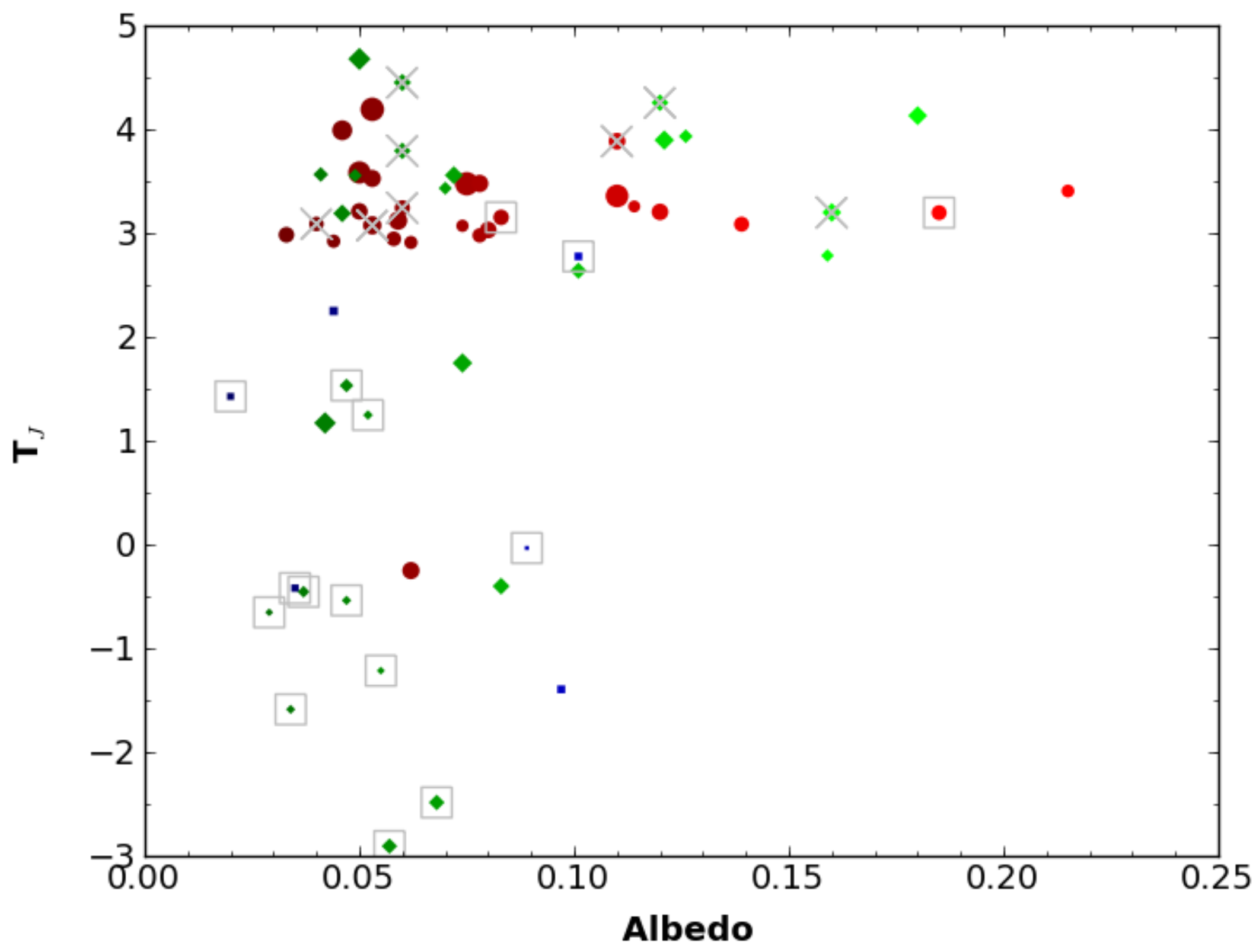

Figure 3: $T_J$, the Tisserand parameter (relative to Jupiter) versus the derived albedos. The symbols are as in Figure 1, with the log value of the fitted diameter represented by the size of the symbol, and the relative visual-band albedo shown approximately as the darkness of each symbol. Note that the low $T_J$ region of the figure tends to be devoid of the higher-albedo objects, but it is unclear whether this is owing to selection bias of optically-selected objects superimposed on objects discovered in the thermal IR.

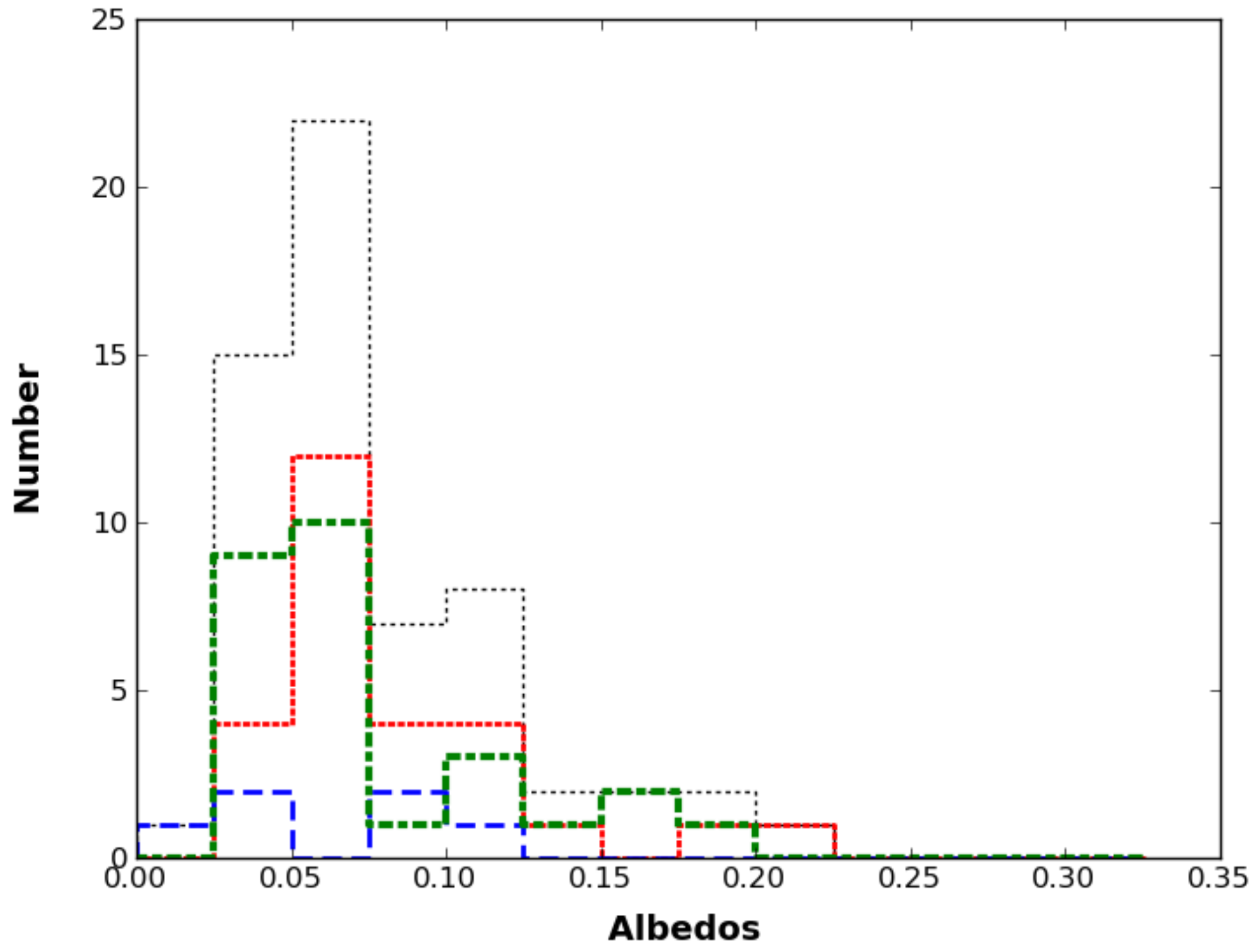

Figure 4: The histogram of albedos for our total sample (black-dotted), Centaurs (red-dotted), SDOs (green-dot-dashed), and the Satyrs (blue-dashed). No particular groupings of albedo with respect to these object classifications are separable by number alone.

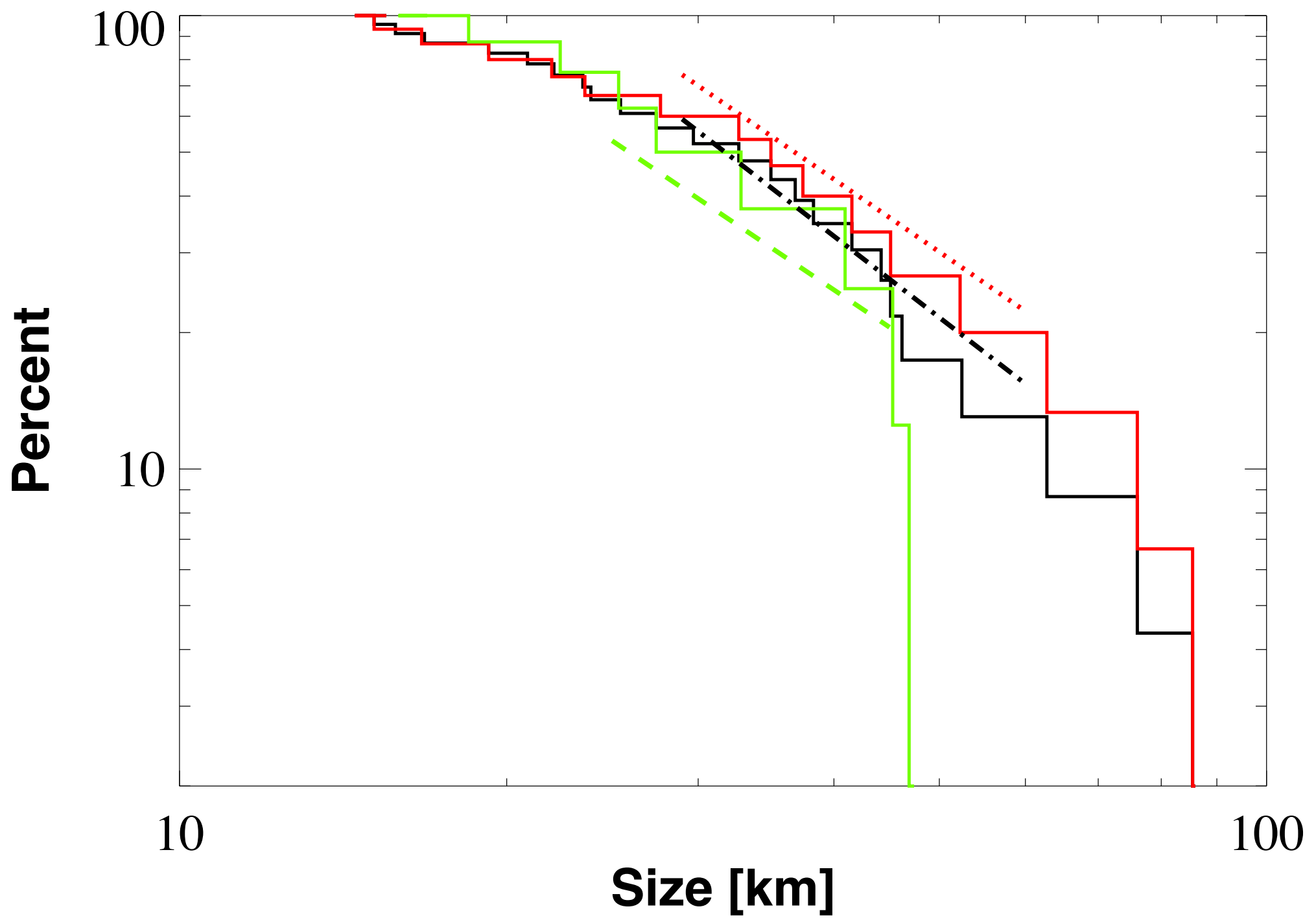

Figure 5: Cumulative size frequency distribution (SFD) of the Centaurs (red) and SDOs (green). The black line indicates the total SFD of the entire sample. The dashed lines indicate the fitted SFD power-law relation for sub-samples of the population less-affected by distance bins (see text), but the power-law distributions of the underlying sample await debaising.

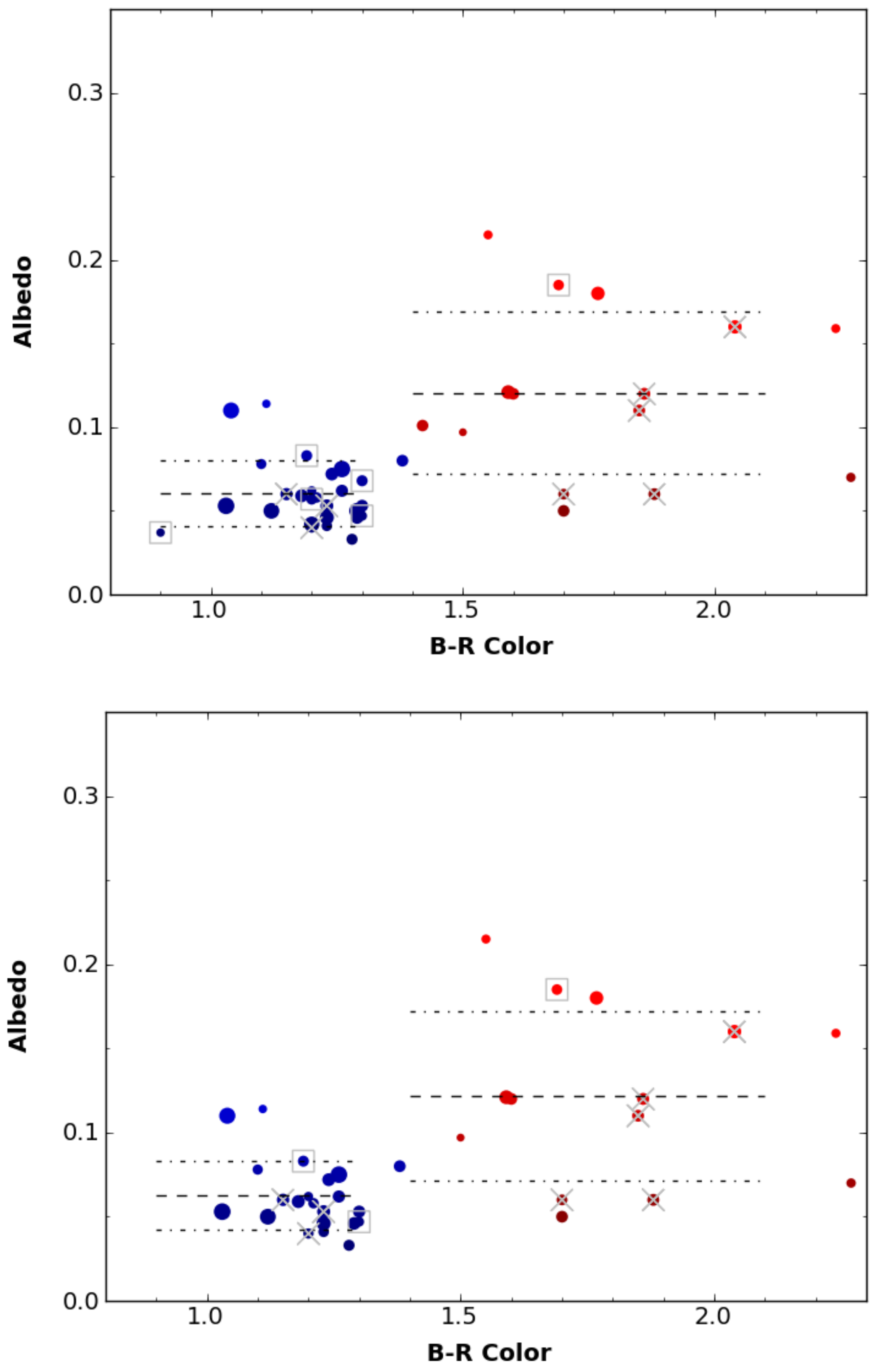

Figure 6: The albedos versus B-R colors of the total sample (top), and Centaur and "Satyr" populations (bottom). The red group (B-R > 1.4) is indicated by the red symbols and the blue (B-R <1.4) by the blue symbols. The mean and standard deviation of each sub-sample are indicated by the dashed and dot-dashed lines, respectively.

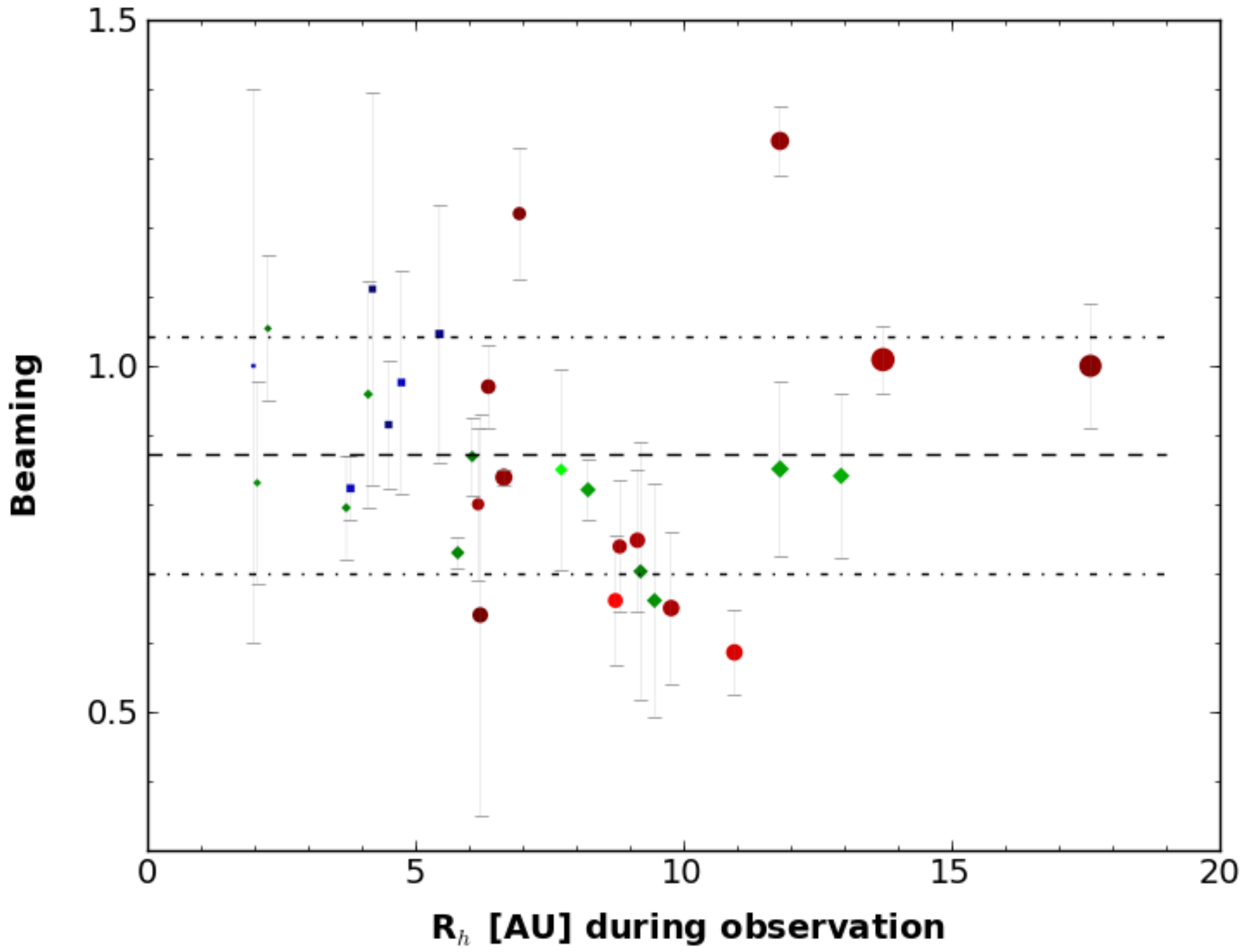

Figure 7: The beaming parameter (η) derived from unconstrained NEATM fits of objects with W3 and W4 signal. The mean and standard deviation of η for the entire sample are indicated by the dashed and dot-dashed lines respectively. The means and distributions of the separate Centaur, SDO, and "Satyr" populations agree with in the total sample

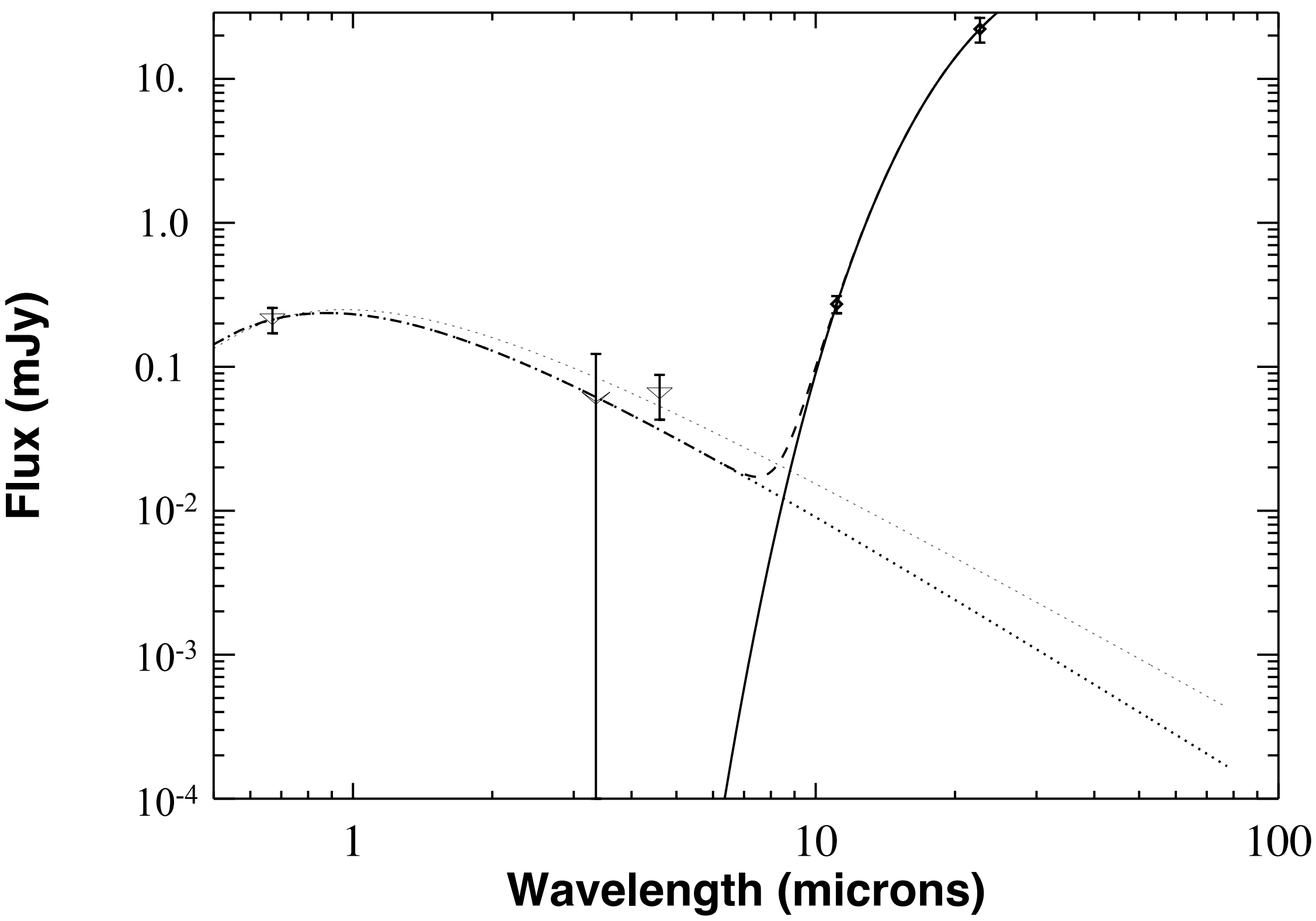

Figure 8: The thermal black body fit to the WISE May 22, 2010 W3 and W4 photometry of the coma and nucleus signal of (2060) Chiron, and the W2 infrared excess, for an 11 arc-second aperture radius. The effective temperature for the thermal fit is 101K. The 1-σ upper bound for Chiron's W1 flux is also shown at 3.4 μm. The dotted lines represent the reflected-light signal based on the observed visual-band brightness of the comet taken on the same UT date and extrapolated using an assumed dust particle-size distribution for a dust particle size distribution power-law of -3.0 (see text; bold dotted line) and with a reddening law based on Jewitt & Meech (1986; faint dotted line). The combined thermal and reflected light signal for the case without the reddening law is indicated by the dashed line.